\documentclass[prd,twocolumn,showpacs,preprintnumbers,nofootinbib,amsmath,amssymb,nobalancelastpage]{revtex4}
\usepackage{graphicx}
\usepackage{bm}
\usepackage{dcolumn}
\usepackage{amsmath}
\usepackage{amssymb}
\usepackage{color}
\usepackage{url}
\usepackage{hyperref}
\usepackage{aas_macros}
\usepackage[normalem]{ulem}

\newcommand{\Msun}{M_\odot}

\newcommand{\mBH}{m_{\rm{BH}}}
\newcommand{\Mmin}{M_{\rm{min}}}
\newcommand{\fNL}{ f_{\rm{NL}}}

\newcommand{\zmax}{z_{\rm{max}}}

\begin{document}

\title{Maximum redshift of  gravitational wave merger events}

\author{Savvas M. Koushiappas}
\email{koushiappas@brown.edu}
\affiliation{Department of Physics, Brown University,  182 Hope St., Providence, Rhode Island 02912, USA}
\affiliation{Institute for Theory and Computation, Harvard University, 60 Garden Street, Cambridge, Massachusetts, 02138, USA}

\author{Abraham Loeb}
\email{loeb@cfa.harvard.edu}
\affiliation{Astronomy Department,  Harvard University, 60 Garden Street, Cambridge, Massachusetts, 02138, USA}

\date{\today}

\begin{abstract}
Future generation of gravitational wave detectors will have the sensitivity to detect gravitational wave events at redshifts far beyond any detectable electromagnetic sources. We show that if the observed event rate is greater than one event per year at redshifts $z \ge 40$, then the probability distribution of primordial density fluctuations must be significantly non-Gaussian or the events originate from primordial black holes. The nature of the excess events can be determined from the redshift distribution of the merger rate. 
\end{abstract}


\maketitle

The discovery of gravitational waves from merging pairs of massive black holes \cite{2015CQGra..32g4001L,2016PhRvL.116f1102A,2016PhRvL.116x1103A,2016PhRvD..93l2003A} has opened a new window to the astrophysics of black holes, their formation, and cosmic evolution. Black holes of stellar masses have been observed with LIGO \cite{2015CQGra..32g4001L,2016PhRvL.116f1102A,2016PhRvL.116x1103A,2016PhRvD..93l2003A} and supermassive black holes in galaxies are expected to be detected by LISA over the next couple of decades \cite{2017arXiv170200786A,2016PhRvD..93b4003K,2006ApJ...639....7K}. The sensitivity of the next generation of ground-based gravitational wave detectors is expected to improve by at least an order of magnitude \cite{2017CQGra..34d4001A}, thus allowing the detection of merging black holes events out to the highest redshifts, potentially exceeding the reach of electromagnetic observations which respond to amplitude squared and not amplitude.

The expected rates of black hole mergers has been calculated based on the number and properties of the few events discovered to-date (see, e.g. \cite{2016PhRvX...6d1015A,2017PhRvL.118l1101A,2017ApJ...840L..24F}). The rate depends on a multitude of factors: black holes must be formed and they must find a way to get close enough so that gravitational waves can take-over as the dominant energy loss mechanism.  The redshift distribution encodes information about the origin of black hole pairs. If black holes originate from massive stellar progenitors then the redshift distribution should relate to the formation, accretion, and cooling of gas in galaxies. If on the other hand the black holes are primordial \cite{1967SvA....10..602Z,1974MNRAS.168..399C,1974A&A....37..225M,1975ApJ...201....1C,1976ApJ...206....8C}, then the redshirt distribution will extend to earlier cosmic times due to primordial binaries \cite{2016PhRvL.116t1301B}. 

A key difference between these two scenarios is that in the case of a baryonic origin, black holes must form out of cold gas, which accreted into a dark matter gravitational potential well, and then cooled to form black hole progenitors. This path follows the abundance of appropriate potential wells. 

In this {\it{letter}} we calculate the  maximum redshift of expected black hole merger events that have baryonic origin in the standard cosmological model. That is, the black holes are formed in galaxies as opposed to primordial black holes, or black holes that are formed in non-standard cosmological scenarios, e.g., cosmologies with a  significant non-Gaussianity in the primordial density fluctuations of the dark matter.  

The significance of this calculation is two-fold: first, it defines a maximum redshift over which baryonic structures can form, and second any detection above the derived bound will signify the presence of either non-Gaussianities that control  the formation of baryonic structures at unexpectedly high redshifts, or that black hole events may be due to primordial black holes.

In the following we make two key assumptions in the derivation of a maximum redshift of baryonic black hole gravitational wave events. First, we conservatively assume that black hole pairs merge instantaneously, i.e., there is no time lag between the formation of black holes, the evolution of the binary and the subsequent sequence of events that leads to a merger. Second, we conservatively assume that all gas accreted in dark matter halos end up in stars that end up in black holes. Realistically, both of these assumptions are vastly optimistic. However, their application guarantees that the derived maximum redshift is indeed a very hard limit and thus any observation that violates this bound will be of enormous scientific significance.

We begin by calculating the number of observed gravitational wave events per year {\it{greater}} than redshift $z$ as the integral of the rate of black hole mergers per redshift interval 
\begin{equation} 
{\cal{N}}(>z)= \int_z^\infty \frac{ d {\cal{R}}}{dz} \, dz,
\label{eq:Ngtz}
\end{equation} 
where $d {\cal{R}}/dz$ is the rate of merger events per redshift interval, 
\begin{eqnarray} 
\label{eq:rate}
\frac{ d {\cal{R}}}{dz} &\equiv&   \int_{\Mmin(z)}^\infty \frac{dN}{dMdV} C_{\rm{NG}}(M,z)   \nonumber \\
&\times&   \frac{\langle \epsilon (M,z) \rangle}{(1+z)} \frac{\dot{M}_{\rm{g}}(M,z)  }{2 \mBH}   \frac{dV}{dz} \, dM. 
\end{eqnarray}
Here,  $dN/dMdV$ is the comoving density of dark matter halos of mass $M$ at redshift $z$, $C_{\rm{NG}}(M,z)$ is a correction to the mass function in the case where non-Gaussianity is present (with $C_{\rm{NG}}(M,z) = 1$ in the standard $\Lambda$CDM cosmology),  $\dot{M}_{\rm{g}}(M,z)$ is the rate of accreted gas in halos of mass $M$ at $z$,  $\langle \epsilon (M,z) \rangle$  is the efficiency of converting gas to black holes of mass $\mBH$, $dV/dz$ is the comoving volume per redshift interval and the $(1+z)$ factor in the denominator is to convert the rest frame rate to the observed rate. 

The integral in Equation~(\ref{eq:rate}) is performed from a minimum halo mass $\Mmin(z)$ to infinity. Throughout the paper we use a  cosmological model with a power spectrum with a spectral index  $n_s = 0.967$, a normalization $\sigma_8 = 0.81$, a present value of the Hubble parameter $H_0 = 70.4 {\rm{km/s/Mpc}}$ and dark matter, baryonic and cosmological constant mass density parameters of  $\Omega_{DM} = 0.226$, $\Omega_b = 0.0455$, and $\Omega_\Lambda = 0.728$, respectively \cite{Planck-Collaboration:2014aa}. 

\begin{figure}
\includegraphics[scale=0.47]{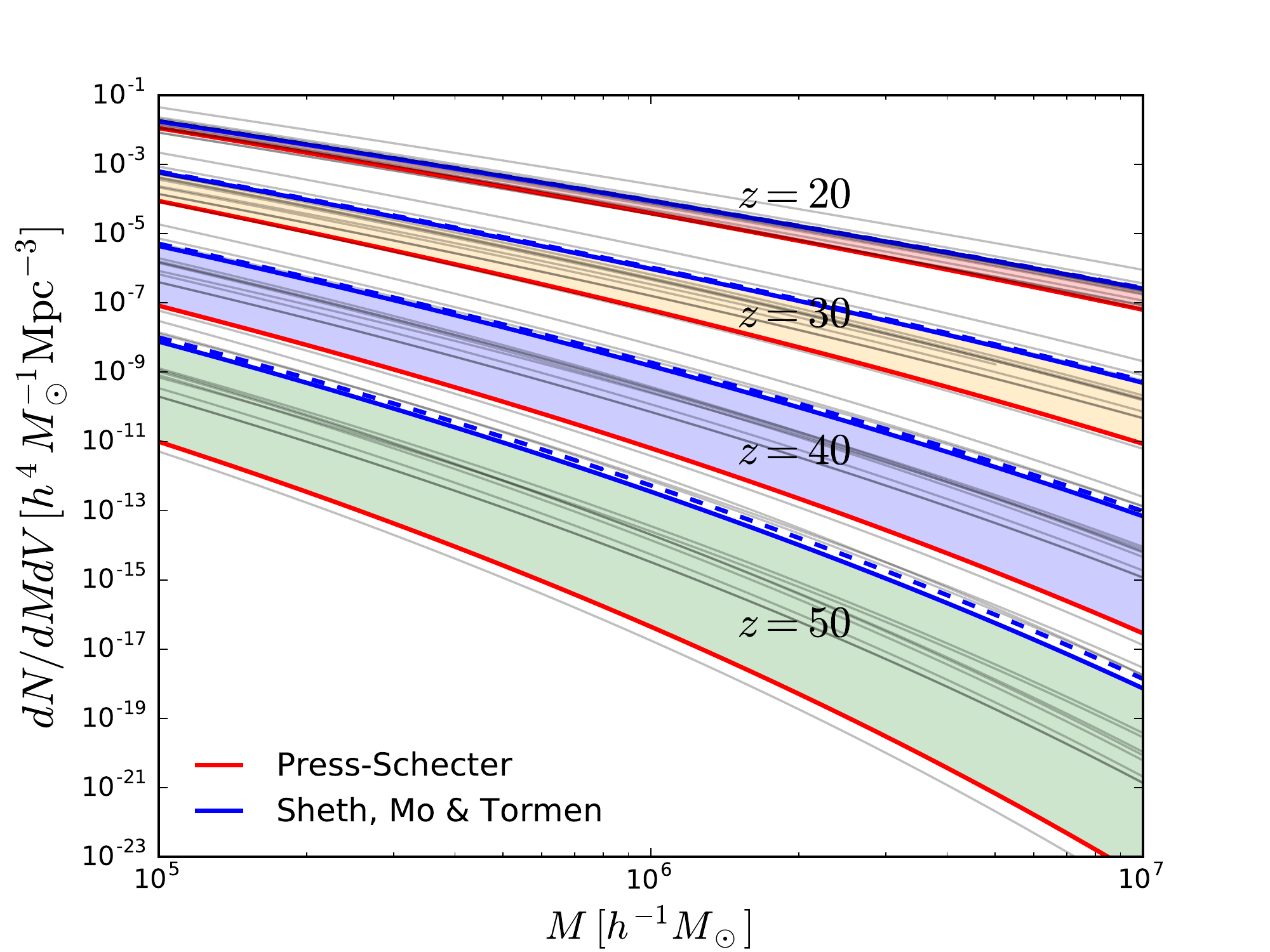}
\caption{\label{fig:fig1} Halo mass functions at $z=$20, 30, 40 \& 50. At each redshift the multiple grey lines correspond to the mass functions derived in Refs. \cite{2012MNRAS.426.2046A, 2013ApJ...770...57B, 2011ApJ...732..122B, 2011MNRAS.410.1911C,2010MNRAS.403.1353C, 2007MNRAS.379.1067P,2003MNRAS.346..565R,2007MNRAS.374....2R,2008ApJ...688..709T,2006ApJ...646..881W,2013MNRAS.433.1230W}. The red line corresponds to the Press-Schecter mass function \cite{1974ApJ...187..425P} while the blue line corresponds to the Sheth, Mo \& Tormen mass function \cite{2001MNRAS.323....1S}. At each redshift, the range of mass function values is bounded roughly by these two analytic mass functions.  The dashed blue lines correspond to the Sheth, Mo \& Tormen mass function with a correction \cite{2000ApJ...541...10M} for a cosmology with a non-Gaussianity parameter $\fNL = 43$ \cite{2016A&A...594A..17P}.  }
\end{figure}

We  next explain how we calculate each of these quantities. The  expected number of gravitational wave events depends strongly on the halo mass function which declines exponentially at high redshifts for Gaussian fluctuations. Figure~\ref{fig:fig1}  shows the mass function at high redshifts from 15 different numerical simulations \cite{2012MNRAS.426.2046A, 2013ApJ...770...57B, 2011ApJ...732..122B, 2011MNRAS.410.1911C,2010MNRAS.403.1353C, 2007MNRAS.379.1067P,2003MNRAS.346..565R,2007MNRAS.374....2R,2008ApJ...688..709T,2006ApJ...646..881W,2013MNRAS.433.1230W,2001MNRAS.323....1S,1974ApJ...187..425P}. The only results that are valid at the high redshifts we consider here are the ones of \cite{2003MNRAS.346..565R,2007MNRAS.374....2R,2013MNRAS.433.1230W}. The halo masses of interest at these high redshifts correspond to extremely rare peaks. The abundance of halos is roughly bounded by two functional forms -- the analytic form of Press-Schecter \cite{1974ApJ...187..425P} (red curve in Figure~\ref{fig:fig1})  gives the lowest number of halos while the ellipsoidal collapse model of Sheth, Mo \& Tormen \cite{2001MNRAS.323....1S} gives the maximum (blue curve in Figure~\ref{fig:fig1}). All other mass functions, including the more realistic results in \cite{2003MNRAS.346..565R,2007MNRAS.374....2R,2013MNRAS.433.1230W} lie in between these two analytic forms.

The presence of a non-Gaussianity in the initial conditions can alter the  abundance of dark matter halos, especially in the exponential tail of the mass function (which is the regime of interest here). We therefore modify the mass function to include such features by assuming that the non-Gaussian mass function is the product of the Gaussian mass function multiplied by a correction factor  \cite{2000ApJ...541...10M} that describes $\fNL$ cosmologies \cite{2009NuPhS.194...22C,2012JCAP...03..002W,2010MNRAS.402..191P,2008JCAP...04..014L} (though it is important to emphasize that $\fNL$ is just one possible parametrization of non-Gaussianities), 
\begin{equation} 
C_{\rm{NG}}(M) =  \left[\frac{\delta_c^2}{6 \Delta} \frac{dS_3}{d \ln \sigma(M)} + \Delta \right]\exp \left( \frac{S_3 \delta_c^3}{ 6 \sigma^2(M)} \right).
\end{equation}
Here, $\Delta \equiv \sqrt{ 1 - \delta_c S_3 / 3}$, $\delta_c = 1.686 \sqrt{a} $, with $a = 0.9$,  and $S_3 = 3.15\times 10^{-4} \fNL / \sigma^{0.838}(M)$. 
The Sheth-Mo-Tormen mass function \cite{2001MNRAS.323....1S}, modified to include the effects of $\fNL$ non-Gaussianities \cite{2016A&A...594A..17P} is shown in Figure~\ref{fig:fig1} as the blue dashed curve. We consider this modified mass function to represent the  maximum abundance of dark matter halos (repeating the calculations for cosmologies with $g_{\rm{NL}}$ or $\tau_{\rm{NL}}$ within the current  limits \cite{2016A&A...594A..17P} leads to smaller effects than the effects from the current uncertainties in $\fNL$).

It is also important to note that the mass function depends on the normalization of the power spectrum; however, the current percent-level uncertainty of $\sigma_8$ is negligible for our purposes. 

\begin{figure}
\includegraphics[scale=0.47]{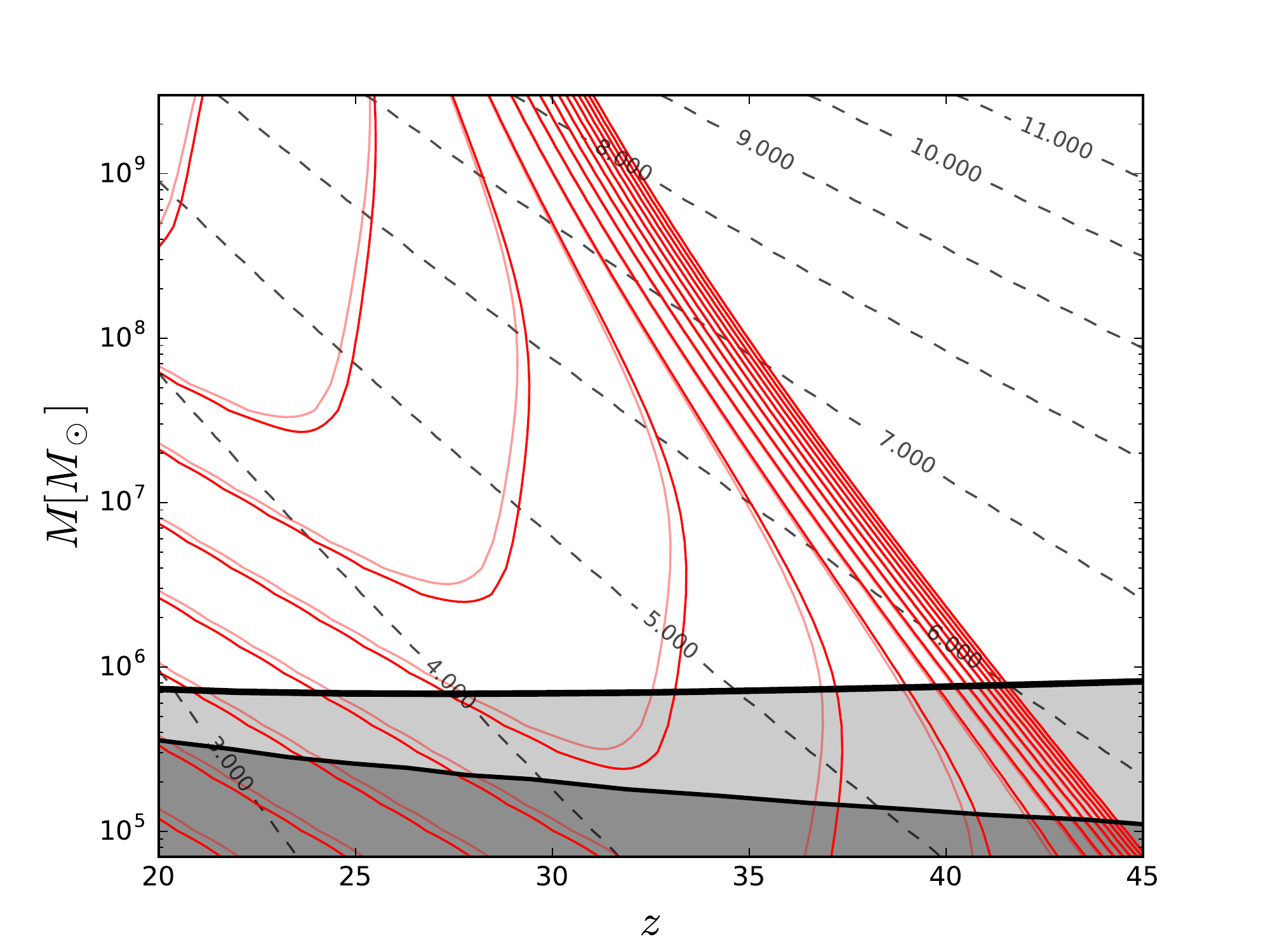}
\caption{\label{fig:fig2}Red contours depict the quantity $\log_{10}[ \langle \epsilon(M,z) \rangle \dot{M}_g/M_\odot {\rm{yr}}^{-1}]$, i.e., the logarithm of the mass accretion rate of gas  that makes black holes in halos of mass $M$ at redshift $z$. The values of the contours are from $10^{-1} - 10^{-10}$ in factors of 10 from left to right. Thick contours correspond to the simulated gas accretion rate of \cite{2016MNRAS.460..417S}, thin contours correspond to the analytic prediction of \cite{2012MNRAS.424..635N}. The dashed grey curves show the number of standard deviations that correspond to the fluctuations of the power spectrum that give rise to halos of mass $M$  at $z$. Black lines correspond to the minimum halo mass for molecular hydrogen cooling. The thick black solid line corresponds to the simulation results of \cite{2012MNRAS.424.1335F} in the case where streaming velocities of gas are  included in the calculation of gas cooling, while the thin solid black line corresponds to the minimum mass when relative motion between gas and dark matter is not considered (see \cite{2012MNRAS.424.1335F}).  By redshift $z \approx 40$ the rate of infalling gas decreases dramatically, while at the same time the minimum mass of a halo that can harbor star formation corresponds to extremely rare density peaks. } 
\end{figure}

The quantity $ \dot{M}_b(M,z))$ represents the rate of gas inflow in halos of mass $M$ at $z$. It has been predicted in simple theoretical grounds \cite{2008MNRAS.388.1792N} and has been measured in hydrodynamical simulations at high redshift \cite{2015MNRAS.454..637G,2016MNRAS.460..417S}. We adopt the maximum gas accretion rate  \cite{2016MNRAS.460..417S},
\begin{equation} 
\dot{M}_g (M,z) \approx 10^{-3} M_\odot {\rm{yr}}^{-1} \left( \frac{M}{10^{6} M_\odot} \right)^{1.127} \left( \frac{1+z }{20} \right)^{\eta},
\label{eq:gasrate}
\end{equation}
where $\eta = 2.5$. Assuming a gas accretion rate as given in  \cite{2015MNRAS.454..637G}  results in a rate that is only slightly smaller  (see Figure \ref{fig:fig2}). If on the other hand we assume that the redshift dependence is steeper (i.e., $\eta > 2.5$) as suggested by high-redshift studies of the growth rate of halos \cite{2005MNRAS.363..393R,2005MNRAS.363..379G}, the gas accretion rate can be higher, however the falloff at high redshift becomes much steeper. Both of these assumptions have negligible effects to the scope of this paper. 

The quantity $\langle \epsilon(M,z) \rangle$ represents the fraction of infalling gas that turns into black holes. In order to compute an extremely conservative upper bound on the number of black holes, we assume that all stars  formed out of infalling gas will be converted to black holes that will merge within the Hubble time at $z$. We assume that this function is
bracketed from above by the ratio of stellar mass to baryon mass in dark matter halos, i.e., $\langle \epsilon (M,z) \rangle = M_{\rm{stellar}} / M \eta$, where $\eta = \Omega_b / \Omega_{DM}$, is the baryon fraction, and we assume the stellar mass as a function of host halo mass and redshift is given by  extrapolating (beyond $z \approx 8$) the results of \cite{2013ApJ...770...57B}.

In Figure \ref{fig:fig2} we show the logarithm of  the product  $\langle \epsilon(M,z) \rangle \dot{M}_g $ in units of $M_\odot {\rm{yr}}^{-1}$, from $10^{-1} - 10^{-15} $ (in declining factors of ten from left to right).  Thin lines correspond to the analytic prediction of \cite{2012MNRAS.424..635N} while thick lines are the numerical results of \cite{2016MNRAS.460..417S}. The contours show that at redshifts $z \le 30$  the gas infall rate increases with mass, and that for a fixed mass it decreases  with increasing redshift at  $z \approx 30$.

At each redshift $z$, we integrate Equation ~\ref{eq:rate} from $\Mmin(z)$ to infinity. The lower mass limit of the integral, $\Mmin(z)$, is the minimum halo mass in which stars can form. This is set by the requirement of the formation of molecular hydrogen  \cite{1997ApJ...474....1T,2001PhR...349..125B}. Recently, it has been argued that the tight coupling of baryons to photons prior to recombination gives rise to a velocity component that becomes important once baryons decouple \cite{2010PhRvD..82h3520T}. Figure \ref{fig:fig2} shows this effect on the minimum mass: the thin solid black curve is the standard case where baryons are assumed to follow dark matter, while the thick solid black curve corresponds to the numerical results of \cite{2012MNRAS.424.1335F} where there is a velocity difference between dark matter and baryons.

Since the minimum mass of molecular hydrogen cooling is roughly constant with redshift, star formation is severely suppressed at increasing redshifts for two reasons: the minimum mass corresponds to extremely rare peaks in the density field (see dashed grey lines in Figure~\ref{fig:fig2} that show the rarity of mass scales as a function of redshift) while at the same time the rate of gas infall decreases rapidly. The combination of these two effects introduces a sharp cutoff to the abundance of stars beyond $z \approx 40$. 

\begin{figure}[t]
\includegraphics[scale=0.47]{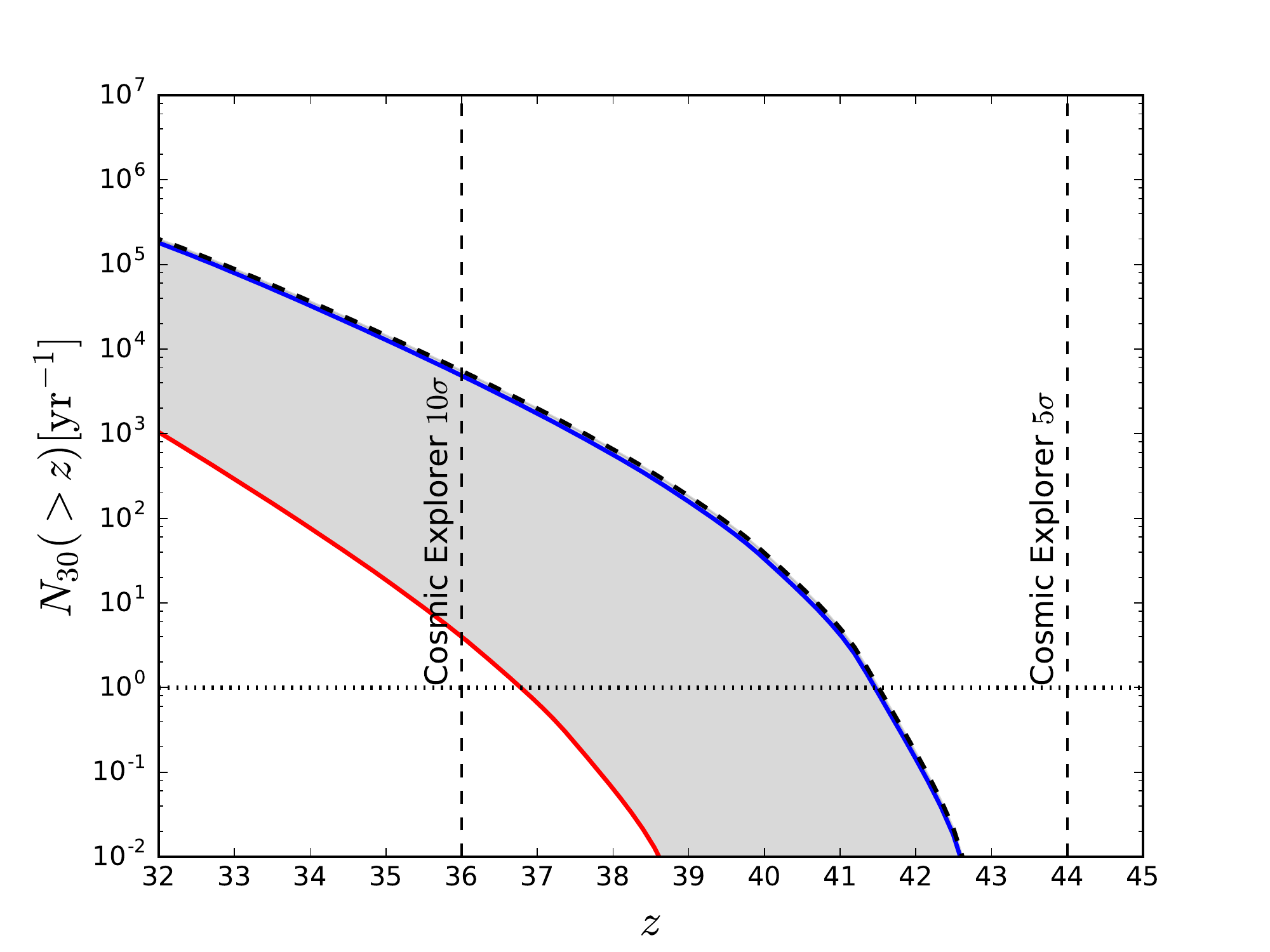}
\caption{\label{fig:fig3}  The number of gravitational wave events of $\mBH = 30 \Msun$ black hole pairs originating from redshifts greater than $z$ (Equation~\ref{eq:Ngtz}) as a function of redshift. The blue curve corresponds to the upper limit on the halo mass function \cite{2001MNRAS.323....1S}, a low value of $\Mmin$ (i.e., ignoring the effects of a relative speed between dark matter and baryons \cite{2012MNRAS.424.1335F}) and a high value of gas accretion \cite{2016MNRAS.460..417S}. The dashed blue curve makes the same assumptions as above, but with a modified mass function that includes a correction  corresponding to non-Gaussianity with $\fNL = 43$ \cite{2016A&A...594A..17P}. The red curve assumes the lower limit on the mass function  \cite{1974ApJ...187..425P}, a large minimum mass (assuming relative velocities between baryons and dark matter \cite{2012MNRAS.424.1335F}) and a low gas accretion rate \cite{2015MNRAS.454..637G}. The shaded area represents everything in between these two extreme cases. The two vertical lines correspond to the $5\sigma$ and $10\sigma$ sensitivity to $\mBH=30\Msun$ black hole pairs with the future gravitational wave detector, Cosmic Explorer \cite{2017CQGra..34d4001A}.
  }
\end{figure}

Integrating the rate of merger events (Equation~\ref{eq:rate}) from redshift $z$ to infinity gives the total number of events per year greater than redshift $z$ (Equation~\ref{eq:Ngtz}). 
Figure~\ref{fig:fig3} shows the result of this calculation. The blue curve corresponds to the maximal mass function \cite{2001MNRAS.323....1S}, a lower $\Mmin$ (i.e., ignoring the suppressing effects of a relative speed between dark matter and baryons \cite{2012MNRAS.424.1335F}), and the maximal value of gas accretion \cite{2016MNRAS.460..417S}. The dashed blue curve makes the same assumptions as above, but with a modified mass function that includes a correction owing to the presence of non-Gaussianity at the current upper bound of $\fNL = 43$ \cite{2016A&A...594A..17P}. The red curve is the opposite of the aforementioned case, where the mass function assumed is at its minimum \cite{1974ApJ...187..425P}, the minimum mass is the largest (including relative velocities between baryons and dark matter \cite{2012MNRAS.424.1335F}) and a low gas accretion rate \cite{2015MNRAS.454..637G}. The  shaded area represents everything in between these two extreme cases.

We define the maximum redshift $\zmax$ such that  the observed event rate is ${\cal{N}}(z = \zmax) = 1  \, {\rm{yr}}^{-1}$. 
We find that the maximum redshift of expected gravitational wave events cannot exceed $\zmax \approx 40$. All assumptions leading to this result are such so that the maximum redshift is maximized: largest abundance of halos (even including current limits on non-Gaussianity), lowest minimum mass for the formation of stars in halos at high redshifts, the assumed gas infall in halos is the maximum measured in numerical simulations, all stars formed in all halos end up in black hole pairs and all black hole pairs merge instantaneously. This confluence of maximizing all assumptions makes the result that the maximum redshift of expected gravitational wave sources of $\zmax \approx 40$ a truly hard bound that cannot be violated unless something very drastically different takes place at high redshifts. 

The aforementioned assumptions can be relaxed and in some cases it is easy to read off the effect on the result (as the vertical axis is a scalable quantity). For example, if all accreted gas ends up in black holes of mass of $\mBH = 10 \Msun$ (instead of $30\Msun$) then the solid curves in Figure~\ref{fig:fig3} simply move up by a factor of 3. If on the other hand only a fraction of  0.1\% of gas ends in black holes of $\mBH = 30\Msun$ then the result of Figure~\ref{fig:fig3} moves down by a factor of $10^{-3}$.

In addition, the assumption of a $\delta-$function mass spectrum of black holes is not realistic. A range of black hole masses is most likely present. The effects of such an assumption have been studied in the context of explaining the current rate of observed black hole merger  events with LIGO \cite{2016PhRvD..94f3530G,2017PhRvD..96b3514C,2017arXiv170907465B,2016PhRvD..94h4013C,2016PhRvL.117i1301M,2016PhRvD..94b3516R,2017PhRvL.118x1101G,2017PhRvD..95j3010K}. In our case, such a black hole mass function will alter the shape of ${\cal{N}}(z )$, but the effect on $\zmax$ is negligible since the factors that give rise to the cutoff remain as discussed earlier (namely the shape of the halo mass function and the decline in gas infall at high redshifts).

The prediction of a maximum redshift for black hole merger events can be tested with  future gravitational wave detectors.  In particular, Cosmic Explorer \cite{2017CQGra..34d4001A} will have the ability to detect events at these very high redshifts. Given the current design capabilities, Cosmic Explorer will be able to detect the merger of 30$\Msun$ black hole pairs at 10$\sigma$ significance out to redshift of $z \approx 36$ and at 5$\sigma$ significance to redshift $z \approx 44$ \cite{2017CQGra..34d4001A}. These two limits are shown as vertical dashed lines in Figure~\ref{fig:fig3}.

Any detection of an event rate greater than once a year from a redshift greater than $\zmax \approx 40$ will have major  implications for cosmology. It would mean that either structure formation is not proceeding in the way that is currently envisioned, or that black hole mergers are  due to some exotic phenomenon. Two such possibilities exist: a strange non-Gaussianity that is not parametrized in terms of $\fNL$ (e.g., decay of cosmic strings \cite{2010AdAst2010E..66R}), or from the merger of primordial black holes \cite{2016PhRvL.116t1301B}. The latter idea has received considerable attention recently in light of the spectacular detection of gravitational waves by LIGO; however at present it seems that other astrophysical constraints make such a possibility less likely \cite{1986ApJ...304....1P,2001ApJ...550L.169A,2007A&A...469..387T,2009MNRAS.396L..11Q,2014ApJ...790..159M,2016arXiv161205644A,2016arXiv161206811A,2016ApJ...824L..31B,2017PhRvL.119d1102K}. Nevertheless, if events with redshifts greater than $\zmax \approx 40$ appear with rates greater than once per year, it may still be possible to disentangle their origin by looking at their redshift distribution as the exact dependence on redshift will be sensitive to the abundance of primordial binaries.

We acknowledge useful discussions with Robert Fisher, Z\'{o}ltan Haiman, Alex Geringer-Sameth, Darren Reed and Kyriakos Vattis. This work was supported by the Black Hole Initiative, which is funded by a grant from the John Templeton Foundation. SMK acknowledges support from DE$-$SC0017993 and the Institute for Theory and Computation at the Harvard-Smithsonian Center for Astrophysics.  We acknowledge usage of the HMFCalc online tool \cite{2013A&C.....3...23M}. 

\bibliography{manuscript}

\end{document}